\documentclass[useAMS,usenatbib]{mn2e}
\usepackage{epsfig,times,graphicx,bm,amssymb,aas_macros}
\newcommand{\beq}{\begin{equation}}
\newcommand{\eeq}{\end{equation}}
\newcommand{\beqa}{\begin{eqnarray}}
\newcommand{\eeqa}{\end{eqnarray}}
\newcommand{\bcom}{}

\title[Failed fossil fields in early-type stars]{Weak magnetic fields in early-type stars: failed fossils}
\author[Jonathan Braithwaite \& Matteo Cantiello]{Jonathan Braithwaite$^1$\thanks{E-mail: jonathan@astro.uni-bonn.de} \& Matteo Cantiello$^2$\\1. Argelander Institut f\"ur Astronomie, Universit\"at Bonn, Auf dem H\"ugel 71, 53121 Bonn, Germany\\2. Kavli Institute for Theoretical Physics, University of California, Santa Barbara, Kohn Hall, CA 93106, USA}
\pagerange{\pageref{firstpage}--\pageref{lastpage}}
\begin{document}\maketitle\label{firstpage}
\begin{abstract} Weak magnetic fields have recently been detected in Vega and Sirius.
Here, we explore the possibility that these fields are the remnants of some field inherited or created during or shortly after star formation and, unlike true fossil fields, are still evolving 
as we observe them. The timescale of this evolution is given in terms of the Alfv\'en timescale and the rotation frequency by $\tau_{\rm evol}\sim\tau_{\rm A}^2 \,\Omega$, which would be comparable to the age of the star. It is shown that it is likely that {\it all} intermediate- and high-mass stars contain fields of at least the strength found so far in Vega and Sirius. Faster rotators are expected to have stronger magnetic fields. Stars may experience an increase in surface field strength during their early main sequence,  but for most of their lives field strength will decrease slowly. The length scale of the magnetic structure on the surface may be small in very young stars but should quickly increase to at least very approximately a fifth of the stellar radius.  
\end{abstract}

\section{Introduction}
\label{sec:intro}

Recent Zeeman polarimetric observations of the main-sequence A stars Vega and Sirius have revealed the presence of weak magnetic fields. In Vega a field of $0.6\pm0.3$ G was found \citep{Lignieres:2009,Petit_etal:2010}, and in Sirius a field of $0.2\pm0.1$ G has been measured \citep{Petit_etal:2011}. The geometry of these fields is poorly constrained at present. 
Also still poorly constrained is any time variability -- although Petit et al.\ note that in Vega, `no significant variability in the field structure is observed over a time span of one year'.
These fields contrast starkly to those measured in the chemically-peculiar Ap stars, which range from about $200$ G to over $30$ kG and do not vary over timescales of at least decades. The field strength distribution amongst the A stars now appears to be bimodal, with a complete lack of field strengths between a few gauss and $200$ G \citep{Auriere:2007}; at least large-scale fields in this range would certainly have been found by now. The situation may well be the same in more massive stars, with kilogauss fields having been measured in a subset of OB stars and the rest possibly having fields below the detection limit \citep{Cantiello:2011}. See \citealt{Donati:2009a} for a review of magnetic fields in relevant kinds of stars.

There are various possible origins of magnetic fields in these stars. In stars with strong, large-scale fields, a `fossil' equilibrium seems the most likely \citep{Cowling:1945,Bra_Spr:2004}. The convective core probably contains a dynamo, but it is very problematic getting this field up to the surface in a sensibly short time \citep{MacGregor:2003}. In massive stars (above about $8M_\odot$) a subsurface iron-ionization-driven convective layer can also host a dynamo, from where there is no difficulty for the resulting magnetic field to reach the surface \citep{Cantiello:2009,Cantiello:2011}; we predict field strengths of (very approximately) $5$ to $300$ G, depending on the mass and age of the star (higher fields in more massive stars and towards the end of the main sequence). These fields could give rise to various observational effects such as line profile variability and discrete absorption components. In intermediate-mass stars such as Vega and Sirius there is a helium-ionization-driven convective layer beneath the surface, from where a dynamo-generated field can float to the surface in the same way -- we explore this in a companion paper (Cantiello \& Braithwaite, in prep.). This could produce fields of approximately the magnitudes observed. However, with the observations done so far, it is probably difficult or impossible to distinguish between this and another hypothesis, namely that these are dynamically-evolving fields, with no need for any ongoing dynamo or other regenerative process. It is this hypothesis that is explored here.

In the next section, the generalities are described of the evolution of a magnetic field in a radiative star in the absence of regenerative processes, before discussing realistic scenarios in section \ref{sec:scenarios}. In section \ref{sec:driving} the effects of convection, meridional circulation and differential rotation are looked at; differences in behaviour between the interior and near the surface are looked at in section \ref{sec:surface}, and the results are summarised in section \ref{sec:discuss}.

\section{Relaxation to equilibrium and timescales}\label{sec:relax}

If one puts an arbitrary magnetic field inside a star it will, in the absence of any driving from differential rotation, convection or meridional circulation, evolve towards an equilibrium. Once it reaches this (stable) equilibrium, it will continue to evolve only on the (very long) Ohmic timescale; this is discussed below. Anyway, we can estimate the timescale over which the non-equilibrium field evolves towards equilibrium, simply by examining the sizes of the various terms in the momentum equation:
\begin{eqnarray}
\frac{{\rm d}{\bf u}}{{\rm d} t} &=& - \frac{1}{\rho}{\bm\nabla}P \;\;\;+\;\; {\bf g} \;\;+\; \frac{1}{4\pi\rho}{\bm\nabla}\times{\bf B}\times{\bf B}\;-\;2{\bm\Omega}\times{\bf u}\;\;\;\nonumber\\
\frac{U}{T} \;& &\;\;\;\;\; \frac{\delta P}{\rho L}\;\;\;\;\;\;\;\;\;\;\;g\;\;\;\;\;\;\;\;\;\;\;\;\;\;\;\;\;\;\frac{B^2}{\rho L}\;\;\;\;\;\;\;\;\;\;\;\;\;\;\;\;\Omega U\nonumber\\
\omega_{\rm flow}^2 & & \;\;\frac{\delta\rho}{\rho}\omega_{\rm sound}^2\;\;\;\;\;\omega_{\rm ff}^2 \;\;\;\;\;\;\;\;\;\;\;\;\;\;\;\;\;\omega_{\rm A}^2\;\;\;\;\;\;\;\;\;\;\;\;\;\Omega\,\omega_{\rm flow}  
\end{eqnarray}
where $U$, $T$ and $L$ are the characteristic velocity, timescale and length scale of the flow -- whereby the general flow relation $U\sim L/T$ is assumed -- and $\omega_{\rm flow}\sim 1/T$ is the inverse timescale of the evolution, $\omega_{\rm sound}=c_{\rm s}/L$ is the inverse sound-crossing timescale, $\omega_{\rm ff}=\sqrt{g/L}$ is the inverse freefall timescale, $\omega_{\rm A}\sim v_{\rm A}/L$ is the Alfv\'en frequency, and $\Omega$ is the angular velocity of the star's rotation. Note that the gravity term can be made to include the centrifugal force, which we therefore do not need to consider separately.

A star without a magnetic field will adjust on the freefall timescale into a spheroidal equilibrium where the pressure and gravity/centrifugal terms balance (although of course any oscillations have to be damped). If an arbitrary magnetic field is then added, the Lorentz force will give rise to motion on spheroidal shells, i.e. perpendicular to effective gravity. 
 If the rotation of the star is slow -- in the sense that $\Omega\ll\omega_{\rm A}$ -- the Lorentz force will be balanced by inertia and we have $\omega_{\rm flow}\sim\omega_{\rm A}$, i.e. we evolve towards equilibrium on a Alfv\'en timescale. This has been confirmed in numerical simulations \citep{Bra_Spr:2004,Braithwaite:2006}. However, if the star is rotating fast -- so that $\Omega\gg\omega_{\rm A}$ -- the inertia term is small and the Lorentz force is balanced instead by the Coriolis force, which gives $\omega_{\rm flow}\sim\omega_{\rm A}^2/\Omega$. Obviously this reasoning ignores geometry, so it is reassuring that precisely this timescale is found for the growth of various MHD instabilities, by e.g.\ \citet{Pitts:1985} as well as in simulations of equilibrium formation (in a forthcoming publication). Thus in the slow and fast rotating regimes we have the following evolution timescales $\tau_{\rm evol}=\omega_{\rm flow}^{-1}$:
\begin{eqnarray}\label{eq:timescales1}
\omega_{\rm A}\gg& &\!\!\!\!\!\!\!\!\!\!\!\!\!\!\!\!\Omega: \nonumber\\ \tau_{\rm evol} \!\!\!&\sim&\!\!\!\! 10^4\,{\rm yr}\!\left(\frac{M/M_\odot}{R/R_\odot}\right)^{\!\!1/2}\!\left(\frac{L}{R}\right)\!\left(\frac{B}{\rm gauss}\right)^{\!-1}\\\nonumber \\\label{eq:timescales2}
\omega_{\rm A}\ll& &\!\!\!\!\!\!\!\!\!\!\!\!\!\!\!\!\Omega: \nonumber\\ \tau_{\rm evol} \!\!\!&\sim&\!\!\!\! 2\!\cdot\!10^{11}\,{\rm yr}\left(\frac{M/M_\odot}{R/R_\odot}\right)\!\left(\frac{L}{R}\right)^{\!2}\!\left(\frac{B}{\rm gauss}\right)^{\!-2}\!\left(\frac{P}{\rm day}\right)^{\!-1}.
\end{eqnarray}

The following considerations can help clarify the situation. In the slowly-rotating case, the Lorentz force on each fluid element is directed in some sense towards equilibrium locations, and although they are forced to move on spherical shells (because of gravity and a positive radial entropy gradient) and such that the divergence of their velocities vanishes (because the field is too weak to cause significant gas compression), there is nothing stopping fluid elements from collectively moving directly towards the nearest available equilibrium. To simplify the situation, let us first consider the case of a single particle which experiences an acceleration $r\omega_{\rm A}^2$ towards an equilibrium position at the origin, where $r$ is the distance from the origin. The inverse-time taken to reach the origin is simply $\omega_{\rm A}$. The lone particle then performs oscillations. In a fluid however, each fluid element will meet and interact with others along the way, and via reconnection events and so on the motion is damped. The equilibrium is therefore reached on an inverse-timescale comparable to $\omega_{\rm A}$. With fast rotation a lone particle performs small epicyles of frequency $2\Omega$ superposed on a slower orbit about the origin of frequency $\omega_{\rm A}^2/2\Omega$. Again, in a fluid we expect this motion to be damped by reconnection events such that an equilibrium is reached on an inverse timescale $\omega_{\rm A}^2/2\Omega$, albeit a different equilibrium from that in the slowly-rotating case. And since the motion is on spherical shells, only the component of the Coriolis force perpendicular to gravity is important, so that the effect is greatest on the rotation axis and vanishes in the equatorial plane.
\footnote{Although the whole calculation is already very approximate, this is a justification for using $\Omega$ rather than $2\Omega$ above, since the magnitude of the Coriolis parameter $f=2\Omega \cos\theta$ averages over a spherical shell to $\Omega$.} 
 Therefore one would expect the field on the equatorial plane to be some kind of equilibrium while the rest of the volume is still evolving. This may mean that the field strength is higher nearer the poles.

\section{Evolving and equilibrium scenarios}\label{sec:scenarios}

After several of these dynamic evolution timescales have passed, an equilibrium is reached. A useful concept in understanding this process is that of magnetic helicity, defined as $H\equiv\int {\bf B}\cdot{\bf A}\,{\rm d}V$ where the vector potential is given by ${\bf B}={\bm\nabla}\times{\bf A}$. Helicity is perfectly conserved in the case of infinite conductivity \citep{Woltjer:1958}, and is {\it approximately} conserved in fluids with finite (but high) conductivity. This can be explained in terms of helicity having units of energy times length: While energy is being destroyed on small length scales in reconnection zones, helicity is little affected. This has been confirmed in various laboratory and numerical experiments, e.g.\ by \citet{Chui_Moffat:1995,Zhang_Low:2003,Braithwaite:2010}. So, relaxing towards a stable equilibrium entails evolving towards an energy minimum for the given value of helicity present. We can write $H=L_hE$ where $E$ is the magnetic energy 
 ($\sim B^2R^3$) and $L_h$ is some length scale, and clearly a lower energy state for a given helicity will have a higher $L_h$. This is presumably the reason why the equilibria observed in Ap stars are often simple dipoles with $L_h\sim R$, which represents the lowest possible energy state. Some have somewhat more complex equilibria; this is presumably because a {\it local} energy minimum has been reached with smaller $L_h$. The theory of equilibrium geometries was reviewed and explored by \citet{Braithwaite:2008}. In any case, a relaxation towards equilibrium will necessarily increase $L_h$. If the initial conditions have small $L_h$ then most of the energy will be lost during the path to equilibrium, and as the field weakens the evolution timescale correspondingly increases, so that the total time required to reach an equilibrium depends essentially on the eventual strength of the equilibrium field rather than on the original field strength.

It is worth noting that $L_h$ is not quite the same as the characteristic length scale $L$ introduced in the last section. One can think of the difference by analogy with a collection left-handed and right-handed screws in a box. The screws have a size $L$ (shaped `fat' so that they have the same size $L$ in all directions) an `energy' $E_s$ and helicity $H_s=\pm L E_s$, where the sign depends on the handedness. In total we have $N$ screws, so that the size $R$ of the box is given by $R^3=NL^3$ and so that the total energy is $E=NE_s$. The total helicity $H$ depends on the relative numbers of right- and left-handed screws; if all the screws are of one kind then the total helicity is $H=\pm NLE_s=\pm LE$, where the sign depends on the handedness. If however both handednesses are present, there is cancellation and the helicity is of smaller magnitude -- if we write $H=L_h E$ then $L_h$ is smaller in magnitude than $L$ and may be positive or negative. In the case of a simple dipole-like torus MHD equilibrium with comparable toroidal and poloidal components, we have just one screw and $L=L_h=R$.\footnote{Since helicity can be thought of as the product of two interconnected fluxes -- it has units of flux squared -- in order to make the analogy more accurate we need to assume the screws have a pitch angle of about $45^\circ$. If this is not the case then we must introduce an extra dimensionless parameter $\zeta$, of order unity or smaller, and then the screws have helicity $H_s=\pm \zeta LE_s$. In fact it is possible that dipole equilibria in stars have a much stronger toroidal field than poloidal, equivalent to $\zeta<<1$; see \citet{Braithwaite:2009}.}

 In any case, 
 we can imagine three different scenarios (with intermediate scenarios between them), corresponding to three different sets of initial conditions at the birth of the star:
\begin{enumerate}
\item The star is born with a strong magnetic field and large helicity, which necessarily means that the initial length scale $L_h$ is not too small. An equilibrium is reached in a very short time, and we observe the `fossil' equilibrium field; the dynamic evolution timescale as given by (\ref{eq:timescales1}) or (\ref{eq:timescales2}) is much less than the star's age.
\item The star is born with a strong magnetic field but very small helicity, meaning that the length scale $L_h$ is very small, either because $L$ is also small (remember that $L_h\lesssim L$), or that $L$ is large but the field has a high degree of symmetry (e.g. two equal and opposite screws). The field evolves quickly at first, but since the predestined equilibrium field strength is very low, the field becomes very weak and the evolution timescale very long, so that the equilibrium is not reached within the stellar lifetime. We observe a non-equilibrium field with evolution timescale comparable to the age of the star.
\item The star is born with very weak magnetic field and therefore very small helicity, regardless of the initial length scales $L$ and $L_h$. Right from the beginning, the evolution timescale is longer than the lifetime of the star. We observe a non-equilibrium field which has evolved little from the initial conditions.
\end{enumerate}

We can see from (\ref{eq:timescales1}) and (\ref{eq:timescales2}) that the kilogauss fields observed in some stars (the Ap/Bp stars and more massive stars with large-scale fields) correspond to scenario (i). This conclusion depends of course on our assumption of lack of driving -- easily justified in the case of slowly-rotating Ap stars, some of which have rotation periods of decades, but a bit shakier in the fast rotators -- one can appeal however to the lack of any major difference between the magnetic properties of the slow and fast rotators. Putting the measured Vega and Sirius field strengths into (\ref{eq:timescales2}) and setting $L=R$ gives an evolution timescale longer than the stellar ages, which are  of order $2-5\cdot 10^8$ yr. That points {\it prima facie} towards scenario (iii), but there are several reasons that scenario (ii) seems likelier; all we need is to reduce the timescale estimate in (\ref{eq:timescales2}) by two or three orders of magnitude. A smaller length scale $L$ seems likely, say $L\sim R/5$ would shorten the timescale by one and a half orders of magnitude.
A shorter length scale would also mean that the observed field strengths may be an underestimate, since some of the detail in the small scales is lost to averaging during the observation.  In addition, it seems very likely that the field strength in the middle of the star is greater than that at the surface (see below), which would also reduce the dynamical evolution timescale. 

Another good reason to favour scenario (ii) over (iii) is that it puts weaker demands on any solution to the well-known flux problem in star formation, the problem of reconciling the enormous magnetic flux threading the interstellar medium with the much smaller flux in the stars which are born out of it (see e.g.\ Braithwaite 2012). 
To make even the most strongly magnetised Ap star all but $10^{-3}$ of the flux must be lost; to make Sirius' field less than $10^{-8}$ can be retained. 
And finally, scenario (ii) makes sense if the young star contained a dynamo field powered by the energy in differential rotation (see the next section) which died away.

It would be interesting to see whether there is any correlation between magnetic properties and stellar age -- a correlation could confirm this picture of ongoing {\it dynamic evolution}, since an ongoing {\it dynamo} should produce a field that depends only on the properties of the star at that moment, which depend on age in a different way. The question remains though of why these stars should have so little magnetic helicity; one might think of a solution in terms of a lack of asymmetry in the star formation process, since helicity is in some sense a measure of the asymmetry of a field, or that convection in the protostellar history of the star produced a small length scale $L$. Also required is an explanation of the huge difference in magnetic helicity present in Ap stars and other A stars; it seems likely that the Ap stars suffered some symmetry-breaking event during their formation.

In this sense it is interesting to note that Ap stars show a complete lack of binary companions with periods shorter than about 3 days (Carrier et al. 2002). It could be that Ap stars are created through the merger of short period binaries (Tutukov \& Fedorova 2010), with their magnetic field resulting from some dynamo process active during the merging phase (see e.g. Soker \& Tylenda 2007, Ferrario et al. 2009). The symmetry breaking involved in the violent process of merging two stars could lead to a much higher helicity. The likely large amount of mass lost during the merger process, together with the presence of a magnetic field, could then lead to a rapid spindown of the merger product, explaining why Ap stars are generally slow rotators. The slow rotation of Ap stars has itself long been a major puzzle --  see \citealt{DAngelo:2011b} for recent progress.

In higher-mass stars, the lifetimes are shorter and a dynamically-evolving field should be correspondingly stronger. For instance, in a star of age $3$ Myr, the field strength should be ten times greater than in Vega and Sirius. However as mentioned above, in stars above about $10 M_\odot$ (at solar metallicity) it is likely that subsurface convection can produce a greater field strength than that expected of a dynamically-evolving field. 

\section{Driving}\label{sec:driving}

Now, it is worth looking at the assumption made above that the magnetic field is not affected by convection, meridional circulation and differential rotation. Strong convection is present in the core of the star, which in an A star is only roughly the first $10$\% in radius and therefore an insignificant fraction of the volume. As mentioned above, a dynamo-generated field will tend to rise buoyantly upwards through the radiative zone (see e.g.\ \citealt{MacGregor:2003}), but it seems that it is impossible to reach the surface within the lifetime of the star. There are other convective layers nearer the surface, of which the most promising is a thin helium-ionization layer lying just $1$\% in radius below the photosphere -- this is the subject of the companion paper: Cantiello \& Braithwaite (in prep.). 
 There are one or two more convective zones even closer to and including the photosphere but the energetics are unfavourable \citep{Petit_etal:2011}. These convective layers, if not actually producing the observed field, will in general however have some effect on a field of the kind described in this paper, since the magnetic field will probably not be strong enough to actually suppress the convection (unlike perhaps in Ap stars). Unfortunately the interaction of convection with an imposed large-scale magnetic field is poorly understood, but it seems plausible that small-scale diffusive processes allow the two to co-exist. See e.g.\ \citet{Moss:1969}.

Meridional circulation and an associated differential rotation of some kind should be present in A stars, as a field of the magnitude observed in Sirius and Vega will be unable to inhibit it (see below). Also present may be some `primordial' differential rotation left over from formation. The picture of a field evolving dynamically undisturbed by outside influence is therefore a little too simple. However the following statements can be made. Differential rotation on its own can only change the azimuthal component of the field, rather than the poloidal component which is seen at the surface. However, differential rotation can produce some interesting effects. Let us assume the star begins its life with a fair degree of differential rotation and a magnetic field of arbitrary geometry. 

 If the magnetic field is above some certain strength, fulfilling the condition $\omega_{\rm A}^3/\Omega^3 > q^2\eta/3\pi^2\Omega r^2$ then the magnetic field should be able to damp the differential rotation without losing its original geometry and strength (\citealt{Spruit:1999} and refs.\ therein). In an A star with rotation period $0.5$ days, the critical field strength is about $15$ G. The differential rotation is damped on the phase-mixing timescale $\tau_{\rm p}\sim\tau_{\rm Ohmic}^{1/3}\tau_{\rm A}^{2/3}$ where $\tau_{\rm Ohmic}$ (see below) is of order $10^{12}$ yr, so that for a $15$ G field we have $\tau_{\rm p}\sim 10^6$ yr, very short compared to the stellar lifetime and also shorter than the timescale on which an equilibrium is approached, unless the field is much stronger than $15$ G; to be precise, the evolution timescale is shorter than the phase-mixing timescale if the field is greater than about $600(L/R)$ gauss in a star with $P=0.5$ days. 
 The field therefore evolves towards it equilibrium {\it after} the differential rotation has been damped.

 If on the other hand the field strength is below this threshold, the field will suffer `rotational smoothing' whereby its non-axisymmetric component is removed \citep{Raedler:1986}. This takes place over a timescale $\tau_{\rm smooth}\sim\tau_{\rm Ohmic}^{1/3}(1/\Omega)^{2/3}$ \citep{Spruit:1999}. Given that the rotation period is normally a lot shorter than the Alfv\'en timescale, the rotational smoothing timescale is clearly much shorter than the phase mixing timescale, shorter than any other evolution timescales. After this smoothing, the star is still differentially rotating but the angular velocity is constant on field lines. The field then evolves towards its equilibrium, which will obviously affect the rotation law.

 However before either of these two states are reached, an instability in the toroidal field may set in \citep{Tayler:1973}. This could lead to a self-sustaining dynamo mechanism \citep{Spruit:2002,Braithwaite:2006b} whereby the instability converts energy from the toroidal to poloidal parts of the field, and the new poloidal field is wound up by differential rotation to create new toroidal field. Eventually this will destroy the original differential rotation, or at least reduce it to a rather low level at which point the mechanism switches off. At this point, the field will continue to evolve dynamically, the field strength decaying, and after some time has passed it is difficult to recognise a field with this origin. 
 However, there may be some symmetry about the rotation axis.

\section{Effects near the surface}\label{sec:surface}

\begin{figure}\includegraphics[width=0.95\hsize,angle=0]{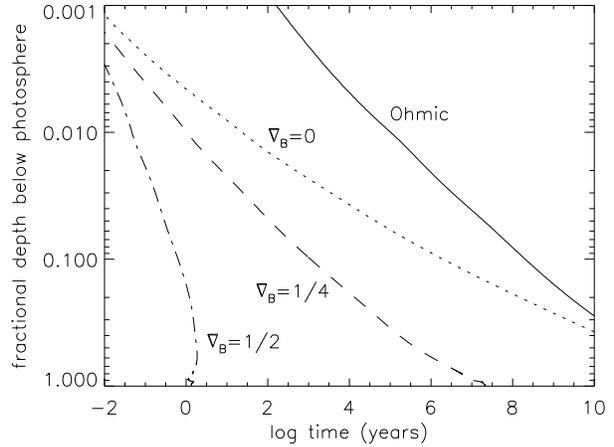}
\caption{The fractional depth (i.e. depth$/R_\ast$) to which dynamic relaxation penetrates into the star, assuming various profiles in the magnetic field of the form $\nabla_B\equiv\partial \ln B/\partial\ln P$, for a $2M_\odot$ model with a field strength of $0.6$ G at the photosphere and a rotation period of $0.5$ days (appropriate for Vega). Also plotted as a function of time is the Ohmic relaxation penetration depth, the depth above which the field should be roughly potential.}
\label{fig:fracdepth}\end{figure}

In the interior of the star the Ohmic timescale is given by $\tau_{\rm Ohmic}\sim L^2 / \eta$ where $L$ is a characteristic length scale and $\eta$ is the magnetic diffusivity. Note that unlike the dynamic evolution timescale, it does not depend on either the field strength or the rotation speed. Putting in the numbers gives $\tau_{\rm Ohmic}\sim (L/R)^2 10^{12}$ yr in an A star and even greater in more massive stars. The field in the bulk of the star therefore evolves purely dynamically unless $L$ is rather small. 
This seems unlikely, given that very small values of $L$ at the surface make the fields practically unobservable. In fact magnetic fields have been detected in both A stars observed so far, despite the fact that cancellation effects preclude the detection of very complex fields using spectropolarimetry. 
However, lower conductivity near the surface of the star will cause the field there to relax to a potential field in a layer beneath the surface. For instance, in a $2M_\odot$ star the potential-field layer should grow to a depth of $\approx0.1R_\ast$ after a time $3\cdot10^8$ yr, calculated by equating this age to $D^2/\eta$ where $D$ is depth and $\eta$ the magnetic diffusivity at that depth. This would mean that the field strength at the surface cannot be much less than at this depth.
 Therefore if the star is born with a very much weaker field at the surface than lower down, which is not implausible if flux is conserved to some extent during the formation of the star, or if the field is created soon after formation by the Tayler-Spruit mechanism (see section \ref{sec:driving}) then the field strength on the surface may actually {\it increase} during the main sequence. This increase would be much faster at the beginning of the main-sequence than later on. In figure \ref{fig:fracdepth} the depth of this potential-field surface is plotted against time (solid line).

Furthermore, since the density is lower near the surface the dynamic timescale is also shorter. Equating an age $3\cdot10^8$ yr to the dynamic evolution timescale $\tau_{\rm A}^2\Omega$ using depth $D$ as the length scale $L$, a field strength (constant with depth) of $0.6$ G and a rotation period of $0.5$ days gives a depth $\approx0.2R_\ast$. The depth of this surface against time is plotted in figure {fig:fracdepth} (dotted line), together with the depths assuming a field-strength-depth relation of the form $\nabla_B\equiv\partial \ln B/\partial \ln P$ with values $1/4$ and $1/2$. The latter is unrealistic but corresponds to $B\propto\rho^{2/3}$ in a radiative star, which one might naively expect from flux conservation during formation, and is shown for completeness.

Therefore in Vega and Sirius we would expect the field to be in approximate dynamic equilibrium down to roughly a depth of $0.2R$ or a little more, while the field is continuously evolving below it. One consequence of this is that the length scale of the field at the surface should not be less than this depth or than the length scale at this depth. Below this depth we can still have the dynamic evolution timescale comparable to the star's age -- we simply need (as mentioned above) a length scale of less than about $0.2R_\ast$ and/or a stronger field. Dynamic evolution however can only reduce the field strength seen at the surface, as new flux cannot be brought upwards since the gas is constrained to move on spherical shells.

We have therefore two competing effects\footnote{Note that in contrast to some other situations near stellar surfaces, magnetic buoyancy is insignificant here since the buoyant velocity scales as $B^2$.} and it is not immediately obvious how the field at the surface should evolve. Certainly its characteristic length scale can only increase, and Sirius and Vega at their current age should have a length scale of $0.1R_\ast$ at the very least, and probably at least $0.2R_\ast$. 

\begin{figure}\includegraphics[width=0.95\hsize,angle=0]{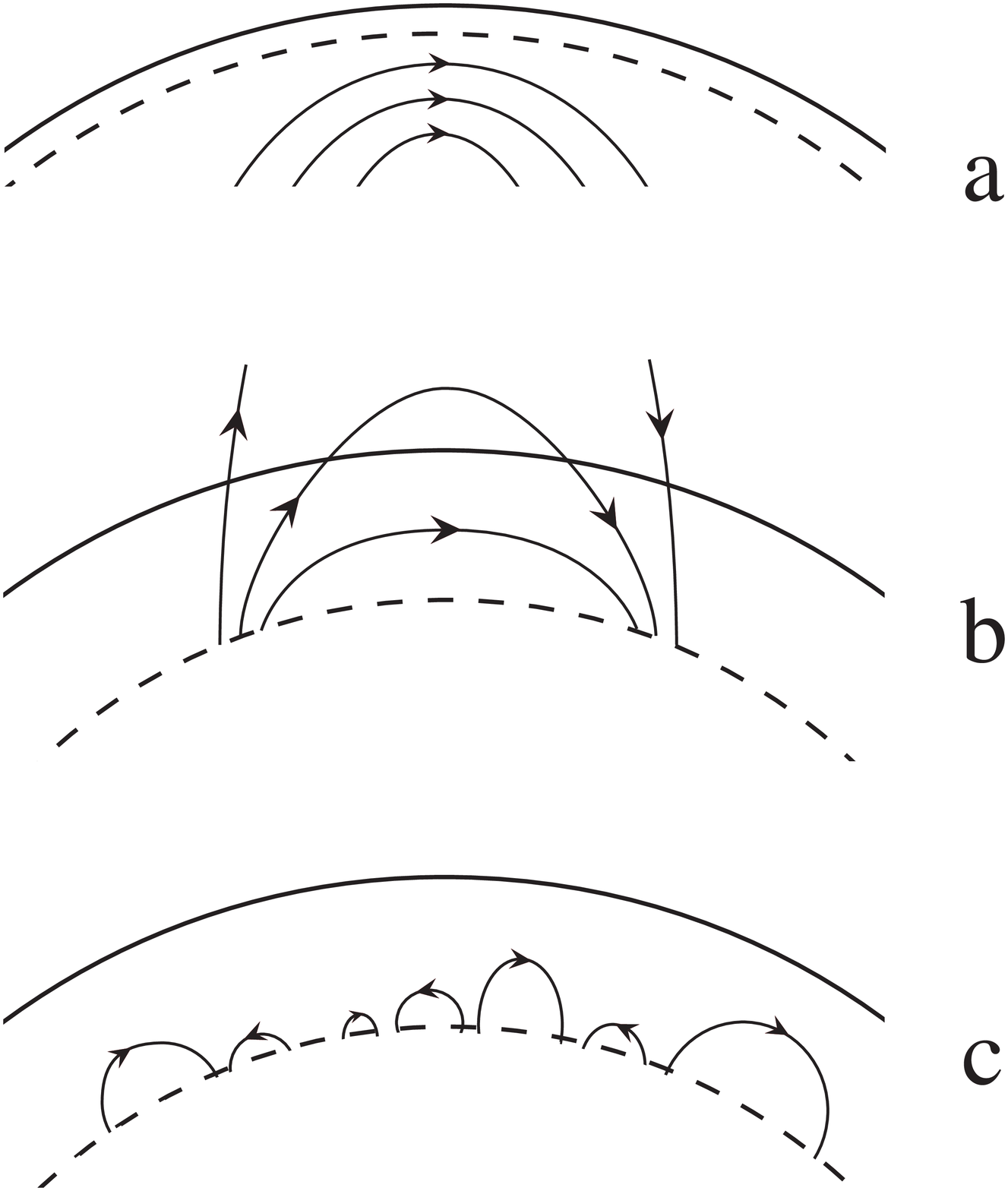}
\caption{Evolution of the fractional depth to which dynamic relaxation penetrates into the star. The limit of this penetration is shown by dashed lines. In the early phase (a) the Ohmic relaxation penetration depth is limited to just a tiny fraction of the stellar envelope. With time Ohmic relaxation penetrates deeper, with the magnetic field present in the stellar interior becoming potential above the dashed line. If the geometry of the magnetic field that is present in the inner layers is not too complex,  the potential field can directly emerge at the surface (b).  Its amplitude and geometry are comparable to the magnetic field present at the Ohmic relaxation penetration depth. On the other hand, if the magnetic field in the stellar envelope has a complex geometry, the potential field will be rather weak at the surface (c).}
\label{fig:bevolution}\end{figure}

Finally, note that in more massive stars the two fractional depths to which the field should be potential and in equilibrium will be somewhat lower, simply because the stars are of limited age.

\section{Conclusions}\label{sec:discuss}

We have presented the hypothesis that the weak magnetic fields found recently on the main-sequence A stars Vega and Sirius do not arise from any ongoing regenerative or dynamo mechanism, but that they are dynamically evolving. This mean that the part of the Lorentz force perpendicular to gravity is balanced not by the pressure-gradient force (as in the case of equilibria in Ap/Bp stars) but by the Coriolis force. The timescale of this evolution is given in terms of the Alfv\'en timescale and rotation angular velocity by $\tau_{\rm A}^2\Omega$.  This timescale would be equal to the ages of the stars. In terms of observables, the tentative predictions of this hypothesis are that the length scale of magnetic structure at the surface of A stars should be no less than (very approximately) a fifth of the stellar radius, but that smaller length scales might be present in more massive or very young stars; that essentially all A stars should have fields of strength at least comparable to the Vega and Sirius fields; that younger stars (more massive or otherwise) should tend to have stronger fields, except that very young stars might experience a field strength increase; and that faster rotators should have stronger fields. Finally, it would be natural to expect some pattern connecting the rotation axis to the magnetic field, i.e. some undetermined kind of symmetry about the rotation axis.\\

{\it Acknowledgements.} The authors would like to thank Fran\c cois Ligni\`eres and Henk Spruit for useful discussions. This research was supported in part by the National Science Foundation under Grant No. NSF PHY05-51164.

\bibliography{../Biblio}

\label{lastpage}

\end{document}